\documentclass{article}

\usepackage{arxiv}

\usepackage[utf8]{inputenc} 
\usepackage[T1]{fontenc}    
\usepackage{url}            
\usepackage{booktabs}       
\usepackage{amsfonts}       
\usepackage{nicefrac}       
\usepackage{microtype}      
\usepackage{graphicx}
\usepackage{doi}
\usepackage{subfigure}

\usepackage{optidef}
\usepackage{amsmath}
\newcommand{\etal}{\textit{et al.~}}
\newcommand{\naive}{na\"ive~}
\usepackage[table,xcdraw]{xcolor}
\usepackage[flushleft]{threeparttable}

\title{Smart Dimming Sunglasses for Photophobia Using Spatial Light Modulator}

\author{
\href{https://orcid.org/0000-0003-0310-0907}{\includegraphics[scale=0.06]{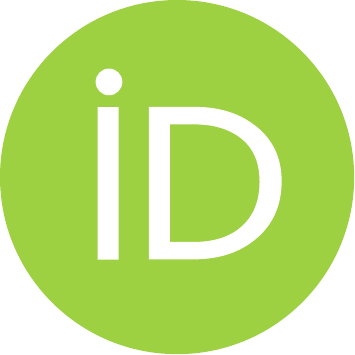}\hspace{1mm}Xiaodan~Hu}\\
	NAIST \\
	\texttt{hu.xiaodan.ht1@is.naist.jp} \\
	\And
	\href{https://orcid.org/0000-0000-0000-00000000-0002-7549-4725}{\includegraphics[scale=0.06]{orcid.pdf}\hspace{1mm}Yan~Zhang} \\
	Shanghai Jiao tong University \\
	\texttt{yan-zh@sjtu.edu.cn} \\ 
        \and
	\textbf{Hideaki~Uchiyama} \\
	NAIST \\
        \texttt{hideaki.uchiyama@is.naist.jp}
        \AND
	Naoya Isoyama \\
	Otsuma Women's University \\
	\texttt{isoyama@otsuma.ac.jp}
        \and
	\textbf{Nobuchika~Sakata} \\
	Ryukoku University \\
	\texttt{sakata@rins.ryukoku.ac.jp} \\
        \and
	\textbf{Kiyoshi~Kiyokawa} \\
	NAIST \\
	\texttt{kiyo@is.naist.jp} \\
}

\renewcommand{\shorttitle}{Smart Dimming Sunglasses for Photophobia Using Spatial Light Modulator}

\hypersetup{
pdftitle={A template for the arxiv style},
pdfsubject={q-bio.NC, q-bio.QM},
pdfauthor={David S.~Hippocampus, Elias D.~Striatum},
pdfkeywords={First keyword, Second keyword, More},
}

\begin{document}
\maketitle

\begin{abstract}
        We present a smart sunglasses system engineered to assist individuals experiencing photophobia, particularly those highly sensitive to light intensity. The system integrates a high dynamic range (HDR) camera and a liquid crystal spatial light modulator (SLM) to dynamically regulate light, adapting to environmental scenes by modifying pixel transmittance through a specialized control algorithm, thereby offering adaptable light management to meet the users' visual needs. Nonetheless, a conventional occlusion mask on the SLM, intended to block incoming light, emerges blurred and insufficient due to a misaligned focal plane. To address the challenge of imprecise light filtering, we introduce an optimization algorithm that meticulously adjusts the light attenuation process, effectively diminishing excessive brightness in targeted areas without adversely impacting regions with acceptable levels of luminance.
\end{abstract}


\section{Introduction} \label{secintro}
	
	Photophobia, or light sensitivity, is a common symptom of various ophthalmic and neurologic conditions, causing discomfort, headaches, or other unpleasant feelings under intense light~\cite{lebensohn1934nature, katz2016diagnosis, burstein2019neurobiology}. This symptom can also be triggered by other factors such as the flicker of fluorescent lights or wavelength-specific light~\cite{vincent1989controlled,good1991use,main2000wavelength}. 
This study focuses primarily on photophobia resulting from heightened light intensity, exemplified by conditions like autism spectrum disorder (ASD)~\cite{Fan2009}. Anomalies related to ASD, such as larger baseline pupil sizes or aberrant pupillary light reflexes, result in an intensified perception of visual brightness~\cite{Fan2009}.

\begin{table*}[]
\centering
\caption{Comparison of tint-changing sunglasses}
\begin{tabular}{cccccc}
\hline
\textbf{\begin{tabular}[c]{@{}c@{}}Tint-changing \\ sunglasses\end{tabular}}  & \textbf{Lens material} & \textbf{Optical sensor} & \textbf{\begin{tabular}[c]{@{}c@{}}Dimming \\ strategy\end{tabular}}  & \textbf{Dimming mode}              & \textbf{\begin{tabular}[c]{@{}c@{}}Response\\ time {[}ms{]}\end{tabular}}         \\ \hline
\rowcolor[HTML]{EFEFEF} 
Photochromic sunglasses~\cite{osterby1991photochromic}          & photochromic           & $\times$               & global                    & binary (on/off)                    & 30,000                            \\
Bhagavathula \etal (2007)~\cite{bhagavathula2007extremely}        & liquid crystal         & pinhole camera          & { \textbf{selective}}  & binary (on/off)                    & { \textbf{50}}            \\
\rowcolor[HTML]{EFEFEF} 
Ma \etal (2008)~\cite{ma2008smart}                  & electrochromic         & $\times$               & global                    & stepwise* adjustable                & 2,000                             \\
Dumas \etal (2012)~\cite{dumas2012fast}               & liquid crystal         & solar sensor            & global                    & {\textbf{flexibly** adjustable}} & {\textbf{50 - 100}}      \\
\rowcolor[HTML]{EFEFEF} 
Chandrasekhar \etal (2014)~\cite{chandrasekhar2014matched}       & electrochromic         & photosensor             & global                    & stepwise adjustable                & 1,000 - 8,000                         \\
Lee \etal (2022)~\cite{lee20203d}                 & electrochromic         & UV sensor               & global                    & stepwise adjustable                & 10,000                          \\
\rowcolor[HTML]{EFEFEF} 
Ctrl eyewear~\cite{AlphaMicron2023}                      & liquid crystal         & light sensor            & global                    & binary (on/off)                    & { \textbf{\textless 100}} \\ \hline
Our system                        & liquid crystal         & camera                  & { \textbf{selective}}  & { \textbf{flexibly adjustable}} & { \textbf{20}}     \\ \hline
\end{tabular}
\begin{tablenotes}
      \small
      \item * Refers to changes made in distinct, predefined levels or steps, not allowing for fine-tuning between these levels.
      \item ** Allows for continuous, seamless changes, enabling precise control without distinct breaks or steps.
    \end{tablenotes}
\label{table}
\end{table*}

Traditional remedies like tinted lenses or sunglasses, while providing some relief~\cite{Clark2017}, come with their own set of challenges. For instance, it is often medically advised against wearing dark or colored glasses indoors, as patients may become dark-adapted and consequently experience increased light sensitivity~\cite{katz2016diagnosis}.

Since the 1960s, Corning and S. K. Deb have pioneered the development of photochromic~\cite{armistead1964photochromic} and electrochromic~\cite{deb1969novel} materials. These materials, when integrated into sunglass lenses, offer a promising solution to the issue of dark adaptation. For instance, they can induce changes in lens color in response to ultraviolet (UV) light~\cite{osterby1991photochromic} and voltage~\cite{ma2008smart, osterholm2015four, lee20203d}, aiding individuals in adapting to different lighting conditions. Nonetheless, photochromic materials suffer from sluggish response times. Although efforts are ongoing to improve their response speeds~\cite{yang2021highly}, achieving a balance between high speed and high contrast remains a significant challenge. Electrochromic materials, while requiring external electricity, are significantly more responsive and sufficiently meet daily requirements. When integrated with light or UV sensors, these materials can be used to develop auto-dimming sunglasses that adjust based on ambient light conditions~\cite{chandrasekhar2014matched,lee20203d}. Yet, the intricacy of redox reactions triggered by variable voltages poses challenges for fine-tuned control over tint or transmittance in electrochromic materials~\cite{rao2022low}.

Liquid crystals, on the other hand, offer rapid and precise control over light transmittance, especially when modulated through advanced sensor setups~\cite{dumas2012fast, AlphaMicron2023}. By segmenting the liquid crystal into multiple independent units, each governed by a distinct electric field, spatial light modulation can be achieved at the pixel level~\cite{hainich2016displays}. This capability is particularly beneficial for photophobic individuals, who are not solely discomforted by glaring sunlight during the day. Indoor fluorescent lighting or the dazzle from oncoming headlights at night can be equally troubling~\cite{Clark2017}. Clearly, devices that allow for selective dimming are more advantageous in high-contrast environments or situations characterized by abrupt shifts in luminosity. For a comprehensive comparison of our system with existing tint-changing sunglasses, see Table~\ref{table}.

Building upon the capabilities of liquid crystals for precise control of transmittance, researchers have delved into the application of transmissive liquid crystal displays (LCDs), a specific type of spatial light modulators (SLMs). These have been employed in a diverse array of applications including vision augmentation~\cite{bhagavathula2007extremely, hiroi2017adaptivisor}, image processing~\cite{Nayar2003,Wetzstein2010}, and optical see-through head-mounted displays (OST-HMDs)\cite{1240696, Itoh2017}. Bhagavathula \etal devised smart glasses that incorporate a transmissive LCD integrated with a complementary metal–oxide–semiconductor (CMOS) and a pinhole panel to form a pinhole camera\cite{bhagavathula2007extremely}. While these glasses are capable of locating and attenuating glare, they offer only rough localization due to the limitations of the pinhole focus, falling short in accurately discerning the size and attenuating the intensity of glare. To enhance accuracy, some researchers have employed cameras as scene detectors and algorithmically calibrated these cameras with the LCD to control pixel transmittance~\cite{Nayar2003, Wetzstein2010, hiroi2017adaptivisor}. While most of these algorithms can provide non-linear dimming—adjusting transmittance based on the brightness of different scenes—none have been designed based on predefined requirements for comfortable vision. Furthermore, achieving accurate selective dimming necessitates the optical focusing of the LCD lens using multiple lenses~\cite{1240696}. In the absence of such an optical arrangement, the dimming becomes ineffective, especially when focusing on distant objects. This ineffectiveness occurs because the occlusion mask, generated by low-transmittance pixels on the LCD, becomes blurred and out-of-focus~\cite{Itoh2017}. A conventional approach to address this challenge is to enlarge the occlusion mask~\cite{Nayar2003, Itoh2017}, as geometric optics dictates that an occluder must be at least as large as the aperture or pupil to fully block a point at infinity. However, this expanded mask inadvertently obscures the surrounding scene, leading to the issue of over-blocking or 'occlusion leak' (as illustrated in the top-left image in Fig.~\ref{fig:result-aperture-method}). While Itoh \etal have proposed a compensation algorithm tailored for OST-HMDs, utilizing virtual images to mitigate this problem effectively~\cite{Itoh2017}, no optimized solution currently exists for eyeglasses.

In this paper, we present a smart sunglasses system designed to overcome these limitations. The system features selective dimming capabilities tailored to predefined requirements for comfortable vision. It incorporates transmissive LCDs and a 12-bit high dynamic range (HDR) scene camera. The higher bit depth of this HDR camera allows for a finer gradation of light intensities, which in turn enables more precise dimming to be achieved by our control algorithm. The scene camera captures the intensity of the surrounding environment, dynamically generating an occlusion mask of low-transmittance pixels on the LCD to filter incoming light. 

Our main contributions include the following: 
\begin{itemize}
\item A compact architecture of smart sunglasses for photophobia.
\item A control algorithm that meets requirements for comfortable vision and operates solely based on input from the scene camera.
\item An eyeglasses-specific optimization strategy aimed at addressing the issue of insufficient dimming due to a blurred occlusion mask.
\end{itemize}

\section{System Architecture and Algorithms} \label{s3}
	In this section, we outline the architecture and key methods of our smart sunglasses system. We cover the benchtop prototype designs in Section~\ref{sec31}, delve into the control algorithm (Section~\ref{sec32})—including its radiometric basics and modulation functions—and conclude with our unique optimization algorithm in Section~\ref{sec33} for precise light attenuation.
	
	\subsection{Benchtop Prototype Designs} \label{sec31}
\begin{figure}
		\centering
		\includegraphics[width=0.8\linewidth]{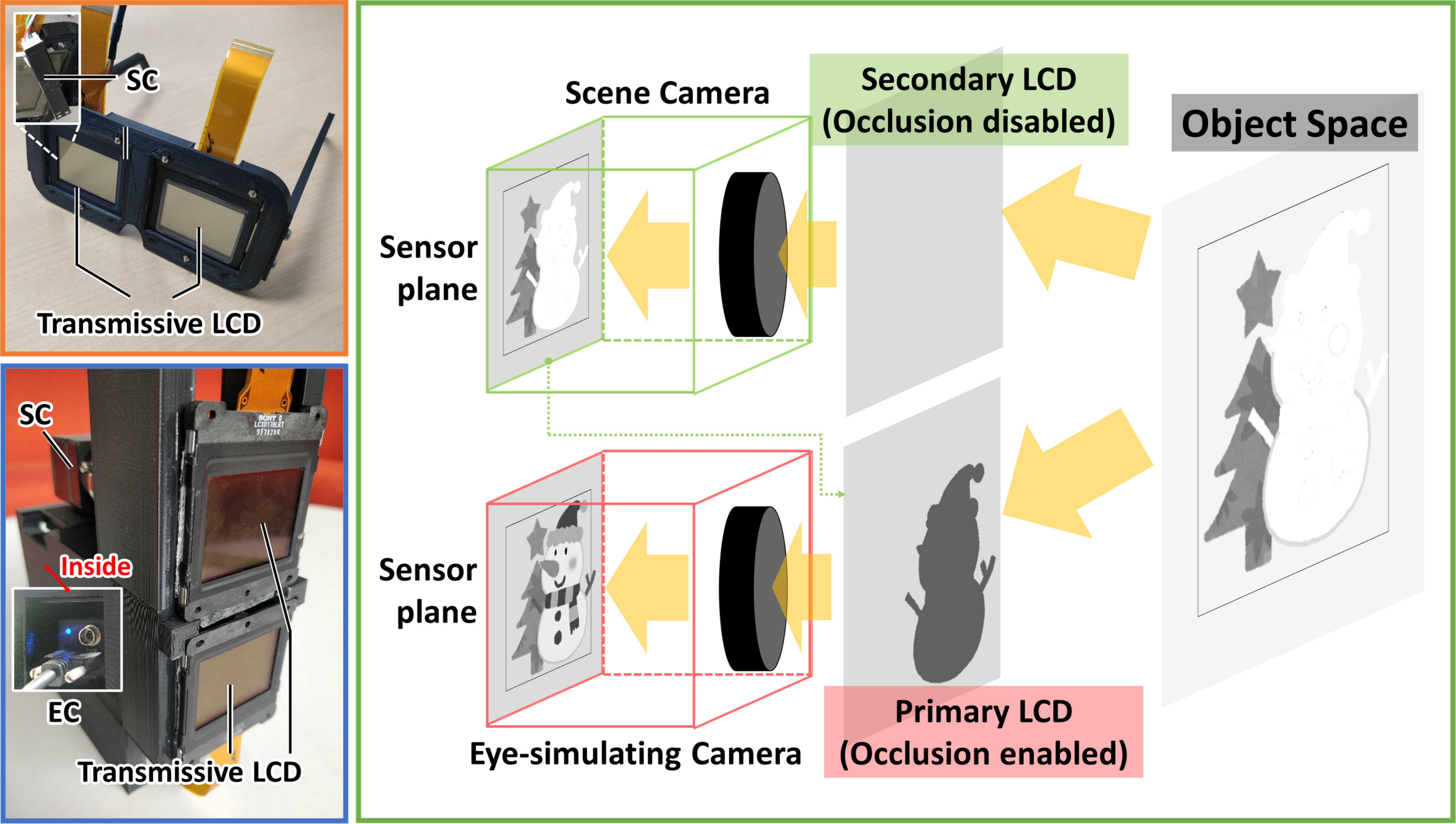}
		\caption{
			(Top left) A proposed smart dimming sunglasses design with an integrated mini scene camera (SC) behind the lens for real-time environmental sensing. 
                (Bottom left) A benchtop prototype setup is used for system evaluation, focusing on light capture and modulation. 
                (Right) The workflow illustrates how the SC first senses the environment through a secondary LCD with uniform transmittance. This data informs the computation of a dynamic occlusion mask, which is then displayed on the primary LCD. Consequently, the eye-simulating camera (EC) perceives the scene with an optimized contrast level on its sensor plane.
		}
		\label{fig:system}
	\end{figure}

	We first built a benchtop prototype (Fig.~\ref{fig:system}(bottom left)) to test and evaluate the system. The benchtop prototype is composed of two parts; a light modulation part and an illuminance detection part (Fig.~\ref{fig:system}(right)). The former consists of a primary transmissive LCD, which displays an occlusion pattern for a human eye, and the latter consists of an HDR scene camera (SC) with a 12-bit depth for capturing a finer gradation of light intensities, as discussed in Section~\ref{secintro}, and a secondary LCD with uniform transmittance that are positioned just above the primary LCD to minimize the parallax. Both LCDs operate in conjunction with a polarizer. To simulate human vision, an eye-simulating camera (EC), which is configured with the same model as the SC, is also incorporated into this prototype setup. We placed the two cameras close together to maximize the field of view (FOV) while avoiding the introduction of a beam splitter.

    Subsequently, we designed a wearable prototype (see Fig.~\ref{fig:system}, top left). In this configuration, LCDs serve directly as lenses, and the scene camera is positioned behind one of them to share a similar structure of the benchtop prototype.

    The workflow is illustrated in Fig.~\ref{fig:system} (right). The SC captures the real environment through the secondary LCD. An occlusion mask is then computed and displayed on the primary LCD, allowing the EC to capture images with balanced contrast.

	\begin{figure}
		\centering
		\subfigure[Optical layout]{
			\includegraphics[width=0.28\linewidth]{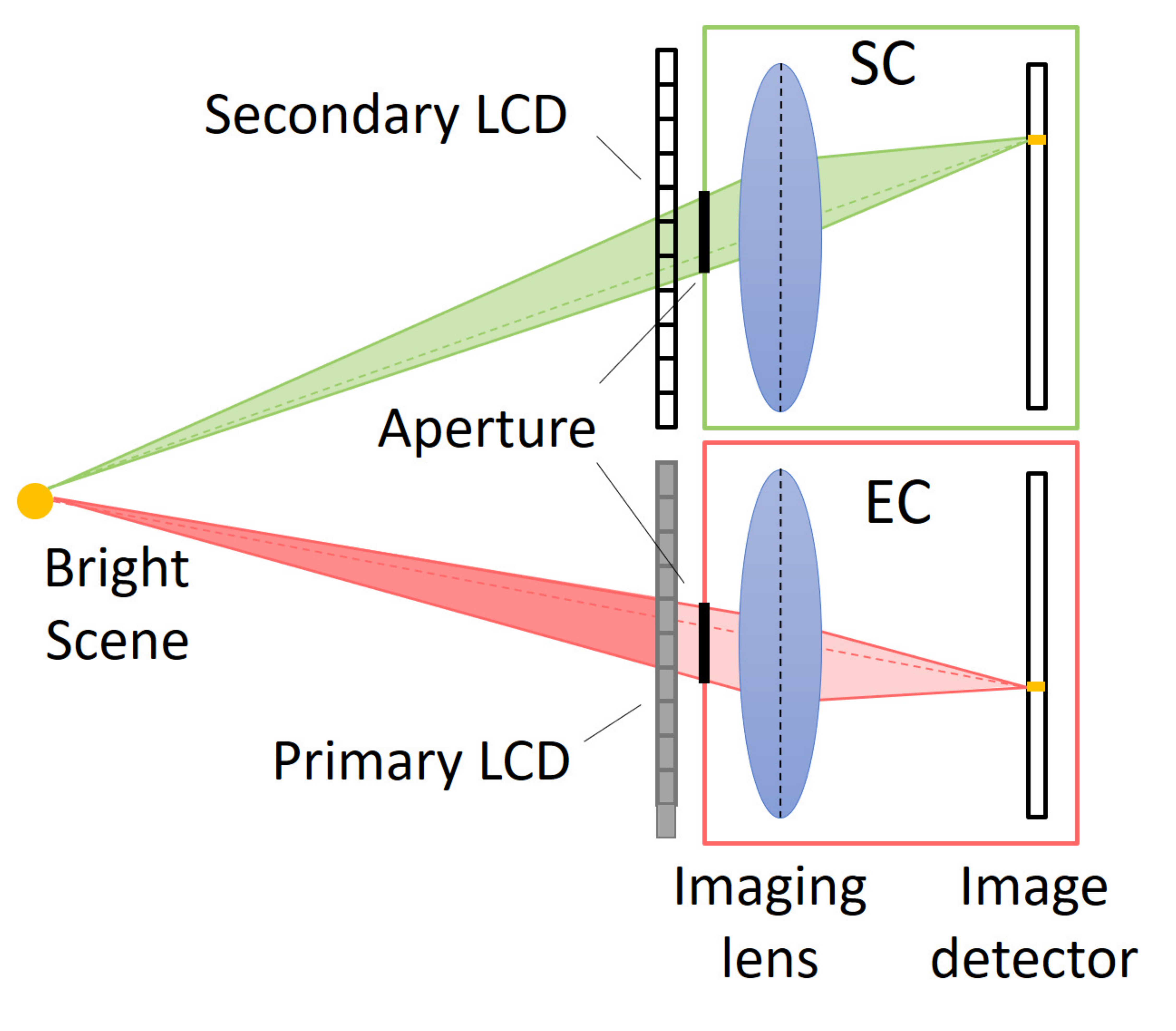}
			\label{fig:layout}
		}
		\subfigure[Schematic figure]{
			\includegraphics[width=0.36\linewidth]{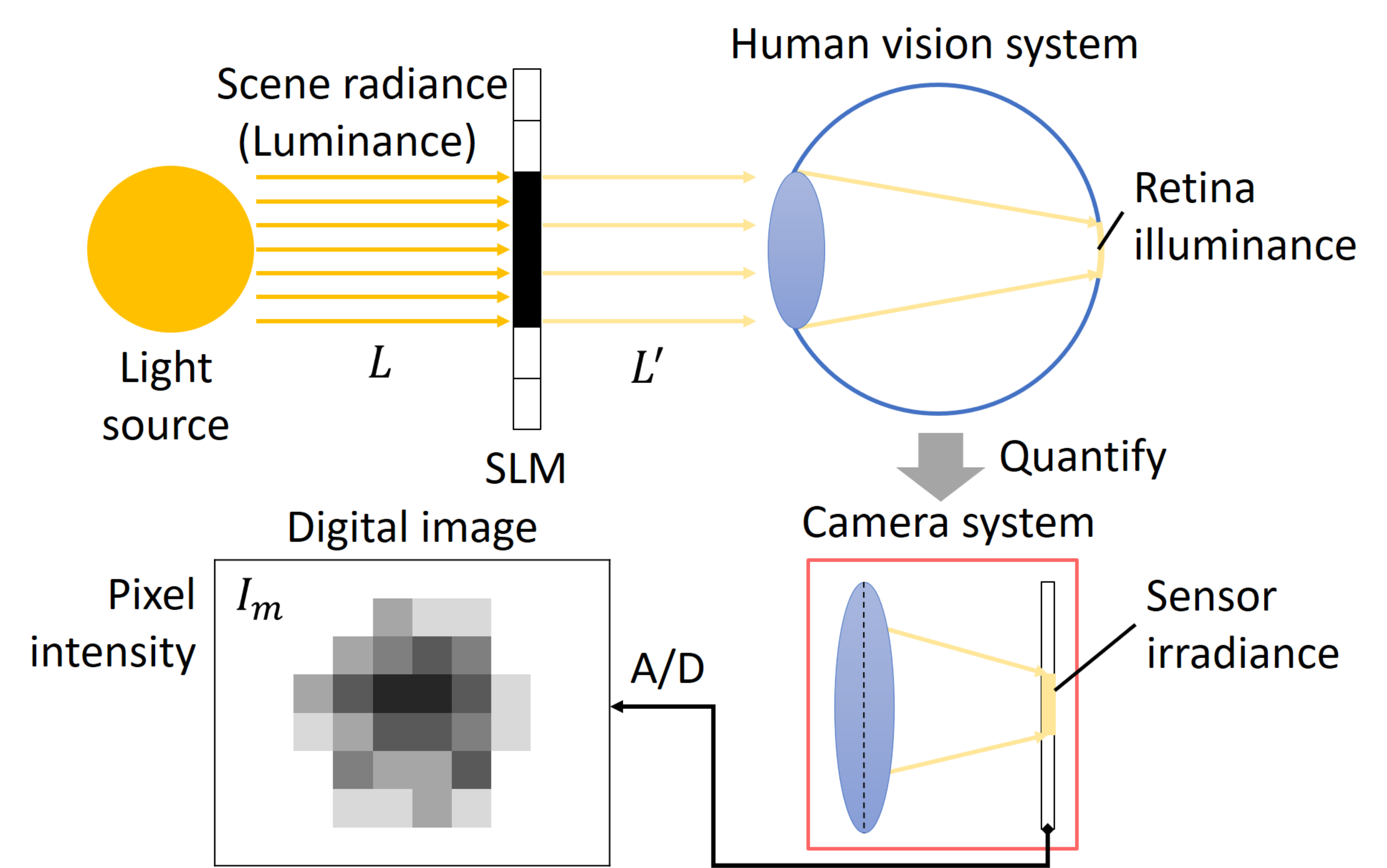}
			\label{fig:radiometry}
		}
		\caption{Optical layout and schematic representation of the proposed system. (a) Two homographic mappings are depicted: one from the primary LCD (for selective dimming) to the image detector of the eye-simulating camera (EC), and the other from the image detector of the scene camera (SC) to the EC. 
                    (b)  The sensed scene radiance, \( L \), undergoes attenuation by the SLM to \( L' \), directing the modified light towards the user's eye. For system calibration and evaluation, the camera's sensor captures the attenuated radiance \( L' \), translating it into sensor irradiance and ultimately into a digital signal characterized by pixel intensity \( I_m \). }
		\label{fig:layout-radiometry}
	\end{figure}

	\subsection{Control Algorithm} \label{sec32}
        \subsubsection{Radiometric Basics}
        
        

        Both human eyes and cameras function by allowing light to pass through a set of lenses, finally reaching a light-sensitive area. In humans, it's the retina; in cameras, it's the image sensor. The sensor converts the incoming light into an electrical signal that is then digitized.
        
        As shown in Fig.~\ref{fig:radiometry}, light attenuates when passing through an SLM. The original radiance \(L\) becomes \(L'\), subject to the transmittance \(T\) of the SLM:
        \[
        L' = TL.
        \]
        
        For our camera-based system, exposure time linearly affects the amount of radiance reaching the image sensor. We assume that the attenuated radiance \( L' \) is linearly related to sensor irradiance. Therefore, the pixel intensities captured by the EC and the SC can be represented as:
        \begin{equation}
		I_{EC}=f_r(L')=f_r(TL), I_{SC}=f_r(T_{max}L),\label{con:2}
	\end{equation}
        where \( I_{EC} \) and \( I_{SC} \) denote the pixel intensities captured by the EC and the SC, respectively. The secondary LCD provides uniform dimming, and its transmittance can be approximated as a constant value $T_{max}$. The response function $f_r$ can be determined through measurements with a luminance meter.
        
\begin{figure}
		\centering
		\subfigure[Attenuation efficiency]{
			\includegraphics[width=0.34\textwidth]{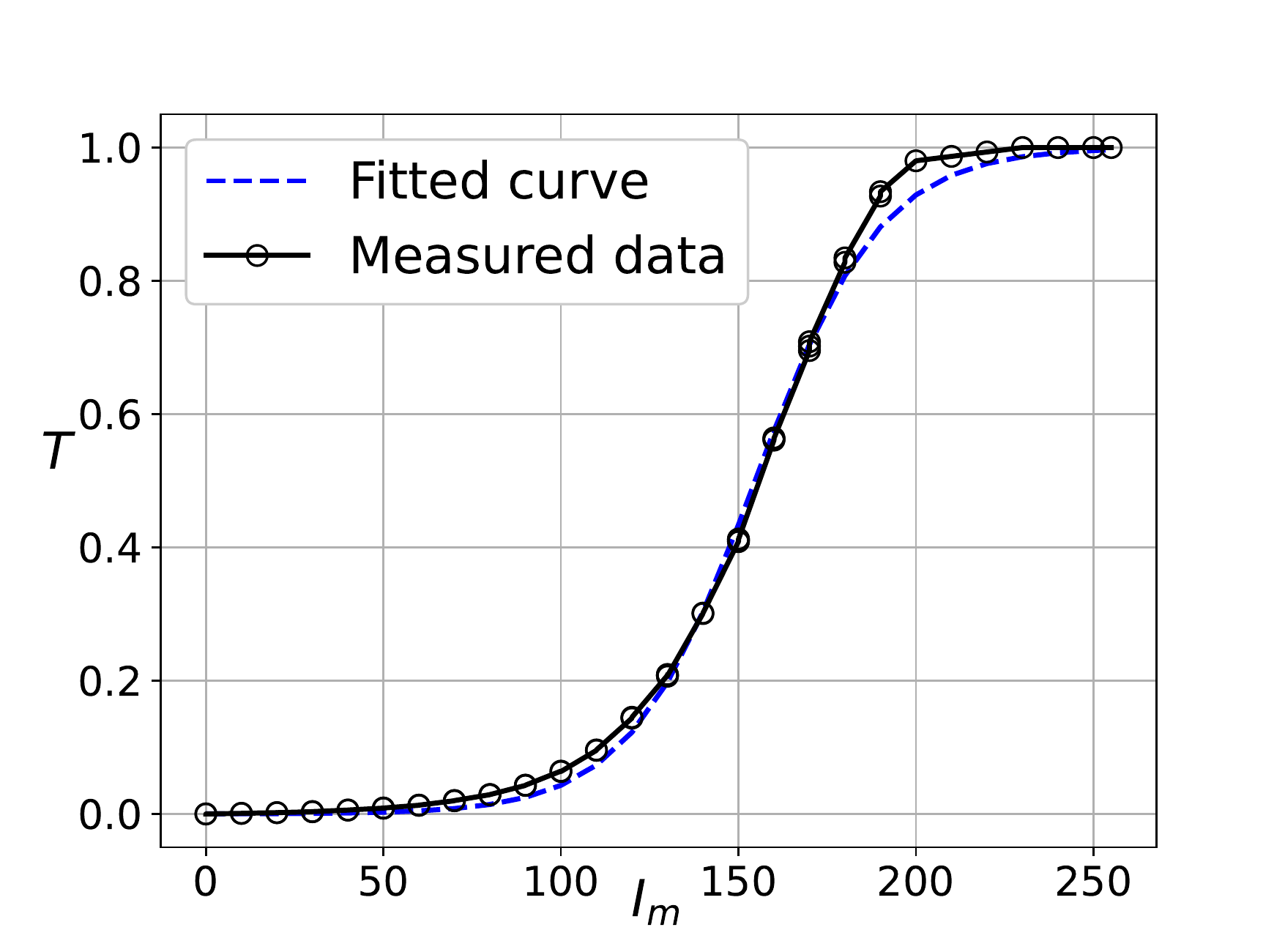}
			\label{fig:effeciency}
		}
		\subfigure[Modulation function]{
			\includegraphics[width=0.315\textwidth]{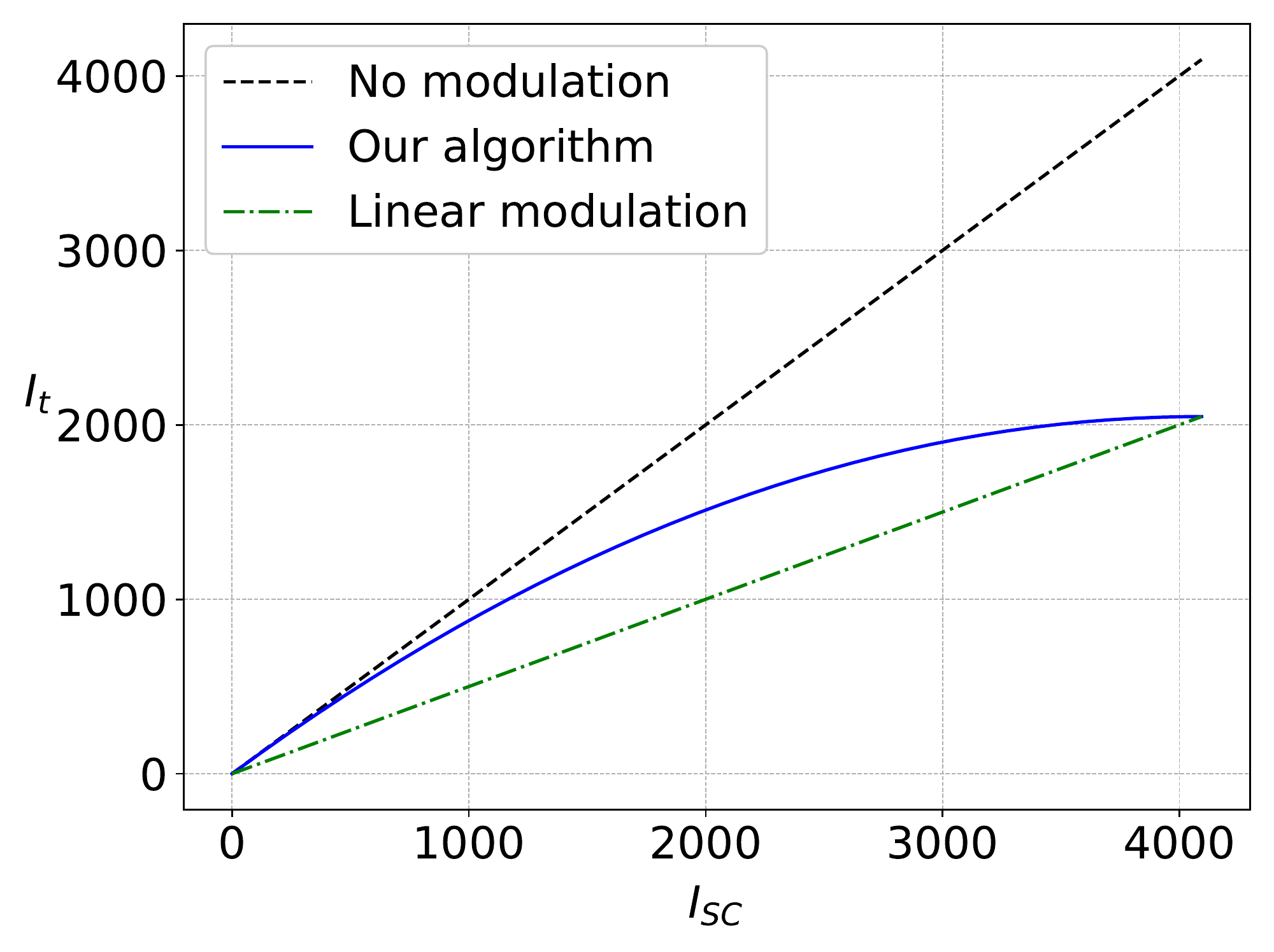}
			\label{fig:function}
		}
		\caption{
			(a) The relationship between the transmittance ($T$) of the LCD we use and the pixel intensity ($I_m$) of the occlusion mask can be fitted by a sigmoid function.
			(b) Modulation functions between original ($I_{SC}$) and target ($I_t$) pixel intensities; no modulation (dashed), linear moduldash-dot), and parabolic modulation (solid).
		}
		\label{fig:effeciency-function}
	\end{figure}

        The transmittance \( T \) is related to the voltage applied to the liquid crystal elements in the LCD. However, as accessing the voltage for each element is impractical, we use pixel intensity \( I_m \) on the LCD as a proxy. 
        We quantity the transmittance $T$ with a certain grayscale value $I_m$ on the LCD by placing a luminance meter after the LCD to measure the attenuated luminance, which is shown in Fig.~\ref{fig:effeciency} (This measurement result depends on the devices, we give the detailed configuration in Section~\ref{sec41}). The relationship can be fitted by a sigmoid function $f_s$ as:
        \begin{equation}
		T=f_s(I_{m}).\label{con:3}
	\end{equation}




	\subsubsection{Modulation Function} \label{sub331}

        In our system, the sole input is the image intensity \(I_{SC}\), derived from the sensor irradiance captured by the SC. The target image intensity on the EC, denoting the perceived brightness for a human observer, is designated as \(I_t\). As per Eq. \eqref{con:2}, \(I_t\) can be formalized as:
        
        \begin{equation}
        I_{t} = f_r(T_{t}L),\label{con:4}
        \end{equation}
        
        where \(T_t\) is the target LCD transmittance for obtaining \(I_t\).
        
        We define the attenuation ratio \(a_t\) as the quotient of \(I_t\) over \(I_{SC}\):
        
        \begin{equation}
        a_{t} = \frac{I_{t}}{I_{SC}},\label{con:5}
        \end{equation}
        
        This ratio represents the relative change in image intensity before ($I_{SC}$) and after modulation ($I_{t}$).
        
        Fig.~\ref{fig:function} delineates the modulation functions. The dashed and dash-dot lines signify the mappings for unmodulated and linearly modulated cases, respectively. Both are uniform modulations that reduce each pixel intensity by a consistent attenuation ratio.
        
        For alleviating the symptoms in photophobic individuals without introducing visual disturbances, we propose the following criteria for computing the occlusion mask:

        \begin{itemize}
		\item High-intensity regions should be comfortably dimmed yet discernible, 
		\item Low-intensity regions should be minimally altered,
		\item Regions with higher (lower) intensity should appear brighter (darker) after modulation.
	\end{itemize}
        
        These criteria are embedded in our modulation function, which is designed to:

        \begin{itemize}
		\item Intersect with the linear-modulation function at the highest intensity,
		\item Be tangent to the no-modulation function at the lowest intensity,
		\item Increase monotonically.
	\end{itemize}
        

        In practice, we have found that a parabolic curve, as shown in Fig.~\ref{fig:function}, satisfies these criteria. This curve not only addresses issues of dark adaptation and glare, but also preserves the natural brightness contrast of the scene, enabling users to easily distinguish between brighter and darker areas. This ensures a more natural and comfortable visual experience with minimal perception of artificial light modulation. Moreover, it's worth noting that the curve and criteria are not fixed and can be customized to cater to different visual needs.

        Hence, using Eq. \eqref{con:2}, \eqref{con:4}, and \eqref{con:5}, we can determine \(T_t\) for any given \(I_t\), and subsequently calculate the pixel intensity \(I_m\) for the occlusion mask using Eq. \eqref{con:3}'s inverse:

        \begin{equation}
        I_m = f_s^{-1}(T_{t}). \label{con:I_m}
        \end{equation}

	\begin{figure}
		\centering
		\includegraphics[width=0.4\textwidth]{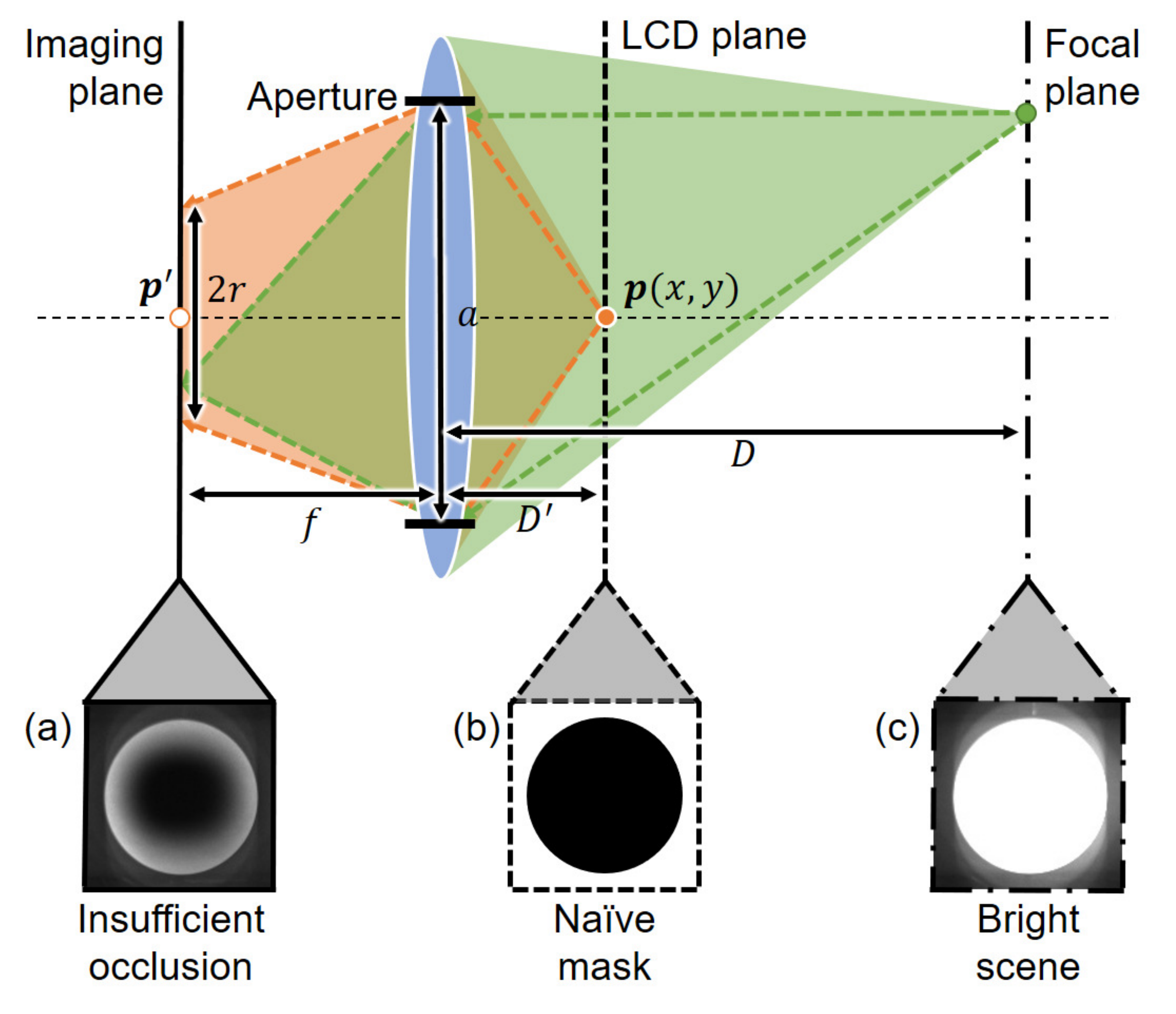}
		\caption{The imaging process of the proposed system. While the LCD panel is not in the focal plane, a pixel $p$ on the LCD plane can be imaged as a blurred disc with $p'$ as the center and $r$ as the radius. With the real bright scene (c), the computed \naive occlusion mask (b) is observed as an insufficient occlusion (a) that allows incoming light from the edge to penetrate the occlusion mask to induce stimuli.}
		\label{fig:focus}
	\end{figure}

        \subsection{Optimization Algorithm} \label{sec33}
	\subsubsection{Out-of-focus Point Spread Function} \label{sub332}

        In real-world scenarios, users often focus on distant scenes. In such cases, the occlusion mask on the near LCD panel appears blurred, a phenomenon described by the out-of-focus point spread function (PSF) \cite{goodman2005introduction}. Fig.~\ref{fig:focus} illustrates this effect.
        
        For a point on the focal plane, the image is in focus. In contrast, pixels on the LCD that deviate from the focal plane manifest as blurred discs in the image. For instance, a pixel \( p \) will be perceived as a blurred disc centered at \( p' \) with a radius \( r \).
        
        If we assume the imaging system follows a linear space-invariant model, we can describe the image formation—neglecting noise and other aberrations—as:
        
        \begin{equation}
        I_{EC}(x,y)=I_m(x,y)\otimes H_{OOF}(x,y),\label{con:7}
        \end{equation}
        
        Here, $H_{OOF}$ represents the PSF of the imaging system, $\otimes$ stands for a convolution operator. Assuming a circular aperture and an LCD plane roughly perpendicular to the optical axis, the PSF \( H_{OOF} \) can be described as:
        
        \begin{equation}
        H_{OOF}(x,y)=\dfrac{1}{\pi r^2} {\rm circ}(\dfrac{\sqrt{x^2+y^2}}{r}).\label{con:8}
        \end{equation}
        
        The radius \( r \) of the blurred disc is determined by:
        
        \begin{equation}
        r = \dfrac{a}{2}\left| 1-\dfrac{D'}{D} \right|,\label{con:9}
        \end{equation}
        
        where \( a \) is the aperture diameter, and \( D \) and \( D' \) are the focal plane and LCD plane depths, respectively. In general, \( D' \ll D \), simplifying Eq. \eqref{con:9} to \( r \approx a/2 \).
        
        As shown in Fig.\ref{fig:focus}, the circular occlusion mask (Fig.\ref{fig:focus}(b)) is computed from the bright scene (Fig.\ref{fig:focus}(c)) detected by the SC. Following a PSF, EC observes a blurred occlusion mask (Fig.\ref{fig:focus}(a)) that is blurred at the edge and leads to insufficient occlusion.

	\subsubsection{Optimization for the Occlusion Mask} \label{sub333}

    As previously mentioned, one conventional approach to enhance occlusion is to expand the occlusion mask by the aperture radius~\cite{Nayar2003}. While effective in blocking light, this method suffers from the issue of occlusion leak, as already discussed. Itoh \etal offered a compensatory algorithm to mitigate this problem~\cite{Itoh2017}, but their solution is not viable for our sunglasses that is without a compensation by the virtual image.
	
	

    In summary, utilizing the original mask leads to inadequate occlusion, while expanding it by the aperture radius results in occlusion leak, primarily because the expansion aims for complete occlusion. Therefore, it becomes crucial to identify an optimal expansion radius that allows the mask to effectively filter light without causing occlusion leak.

    To tackle this challenge,  we formulate a mathematical model aimed at optimizing the occlusion process. Our primary objective is to enable the central pixel to maintain its intensity without excessive degradation, thereby achieving an \textit{effective occlusion}. Simultaneously, the model aims to ensure that these pixels do not leak beyond the intended occlusive area, leveraging the natural defocusing at the mask's edge to minimize perceptibility.
	

    In Eq. \eqref{con:4}, \eqref{con:5}, and \eqref{con:I_m}, we collectively describe how the scene brightness, \(I_{SC}\), impacts the pixel intensity, \(I_m\), displayed on the LCD panel. To illustrate this, let's consider an extreme scenario where the scene resembles an approximately binarized image, as depicted in Fig.~\ref{fig:focus}(c). In this case, only the central bright region exists, surrounded by a dark area. The corresponding occlusion mask to be displayed on the LCD would also be binarized, as shown in Fig.~\ref{fig:focus}(b).

\begin{figure}
		\centering
		\subfigure[]{
			\includegraphics[width=0.18\textwidth]{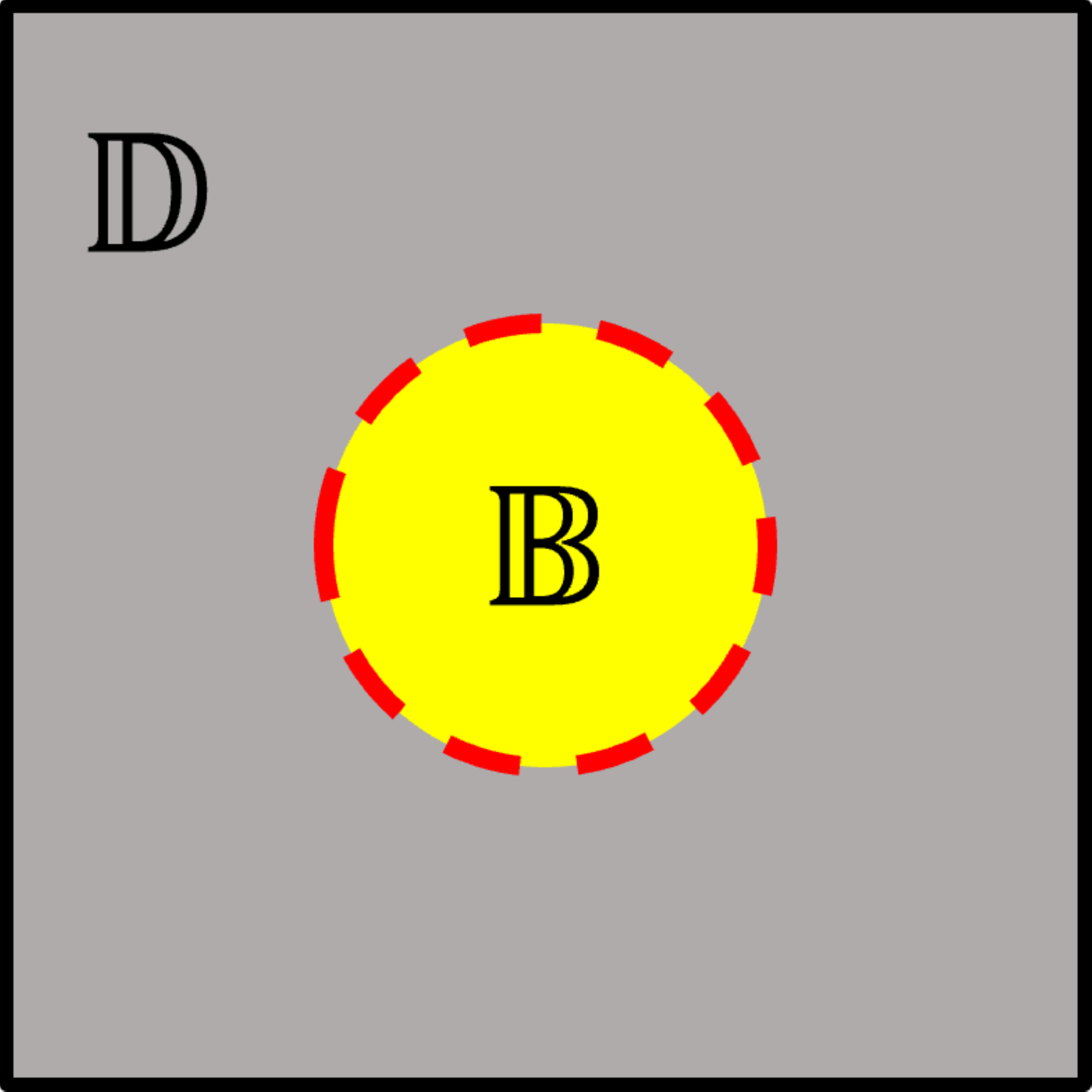}
			\label{fig:effectiveocclusion1}
		}
		\subfigure[]{
			\includegraphics[width=0.18\textwidth]{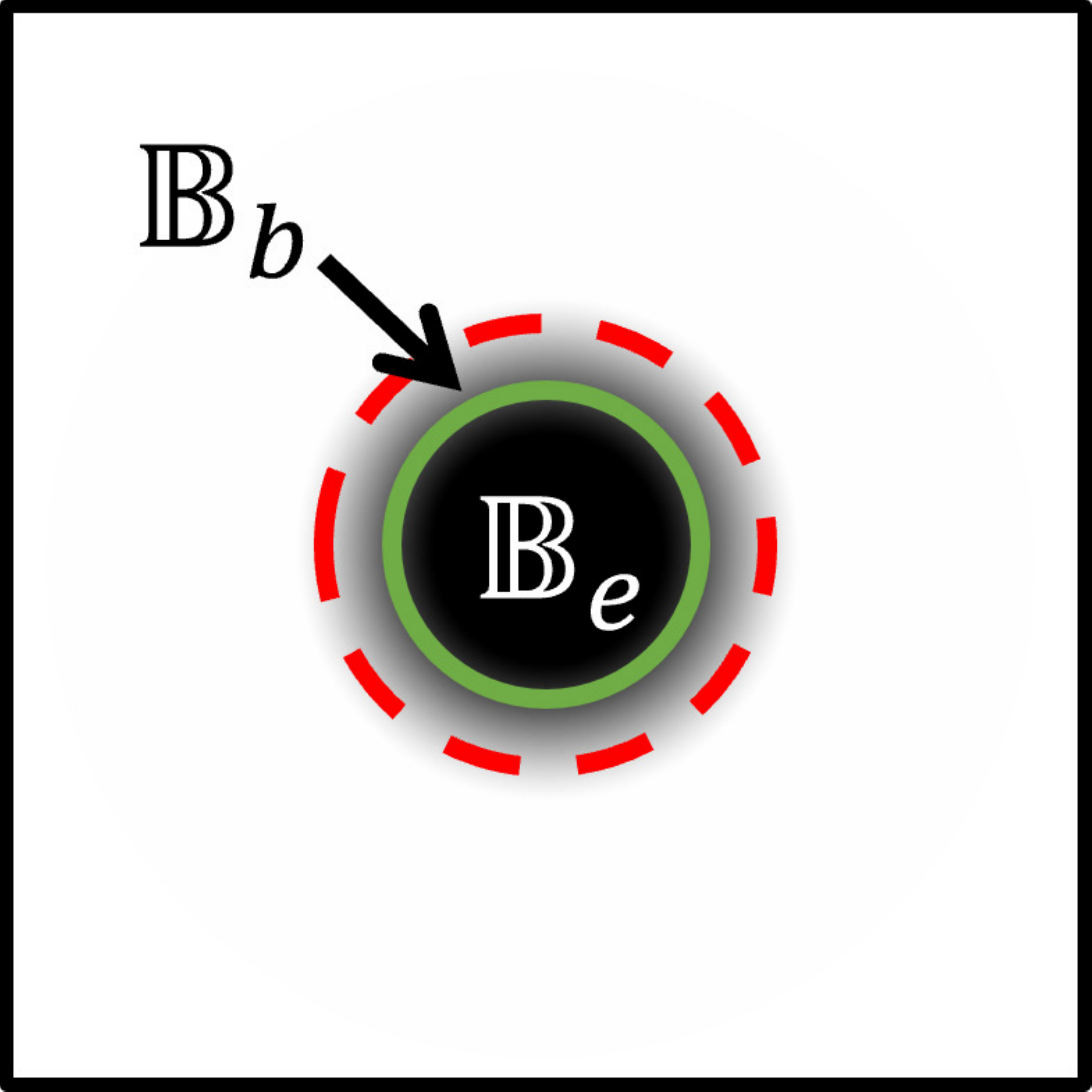}
			\label{fig:effectiveocclusion2}
		}
		\subfigure[]{
			\includegraphics[width=0.18\textwidth]{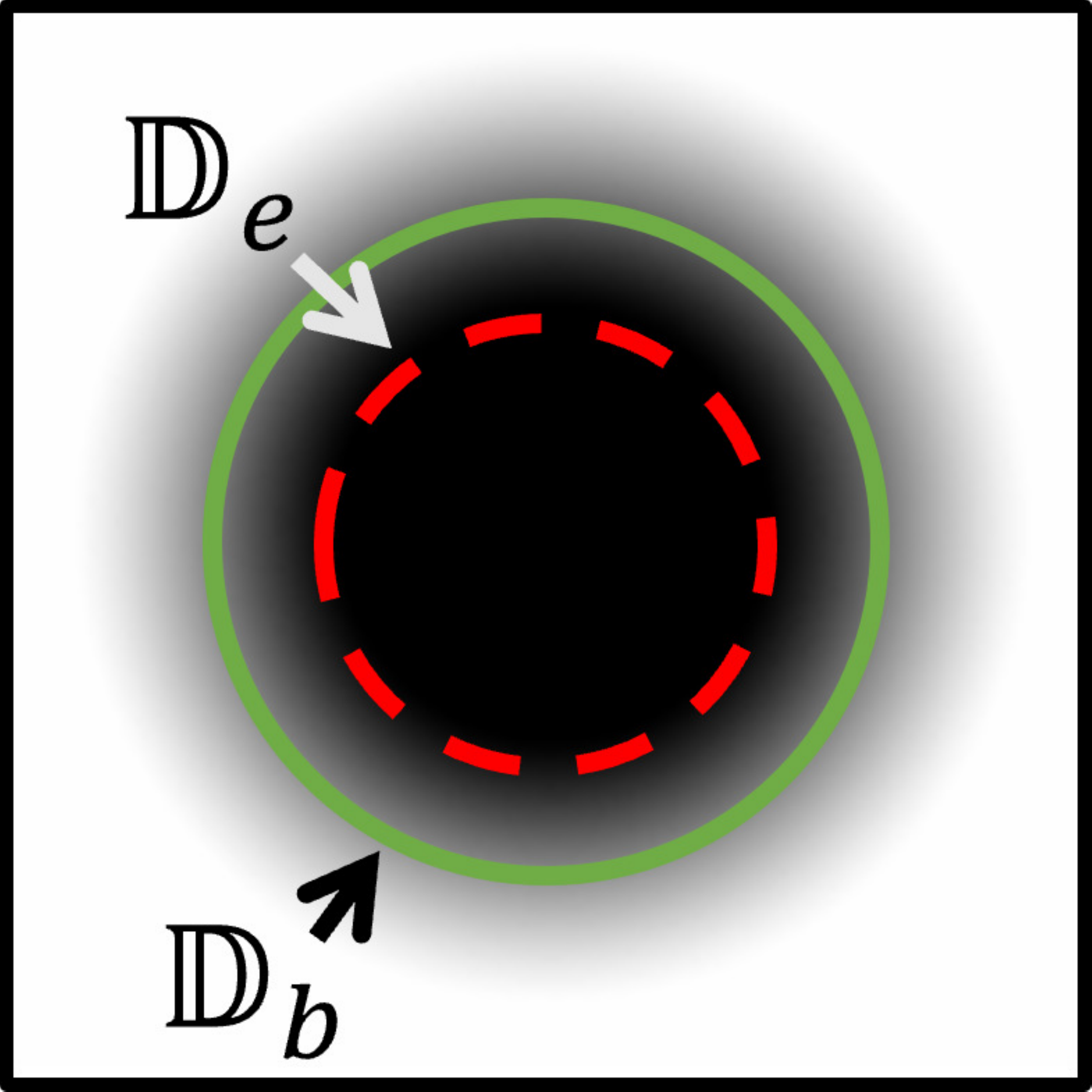}
			\label{fig:effectiveocclusion3}
		}
		\caption{
        Visualization of Eq. \eqref{con:10}-\eqref{con:15}. 
(a) Schematic diagram captured by the scene camera. The bright area inside the red dashed circle represents the set \( \mathbb{B} \), while the darker area outside denotes the set \( \mathbb{D} \). 
(b) Schematic diagram from the eye-simulating camera showing insufficient mask occlusion. The green solid line differentiates the \textit{effective} occlusion from the blurred occlusion. The area within the green solid circle is the set \( \mathbb{B}_e \), and the annulus between the red dashed circle and the green solid circle represents \( \mathbb{B}_b \). 
(c) Schematic from the eye-simulating camera where the mask is overly extensive, resulting in occlusion leak. The set \( \mathbb{D}_e \) is depicted by the annulus between the green solid circle and the red dashed circle, while the blurred occlusion outside the green solid circle is the set \( \mathbb{D}_b \).
		}
		\label{fig:effectiveocclusion}
	\end{figure}

    We define the bright (\(\mathbb{B}\)) and dark (\(\mathbb{D}\)) regions in the image captured by the SC as follows, schematically represented in Fig.~\ref{fig:effectiveocclusion1}:
    \begin{equation}
        \mathbb{B} = \{(x,y)|I_{SC}(x,y) \geq t_{b}\}, \mathbb{D} = \overline{\mathbb{B}},\label{con:10}
    \end{equation}
    where \( t_{b} \) is a threshold for defining the bright region and \( \overline{(\cdot)} \) denotes the complement of a set.

    To accomplish \textit{effective occlusion}, we focus on the intersection of pixels that maintain their intensity within acceptable bounds with the bright region, as shown in Fig.~\ref{fig:effectiveocclusion2}. This set of pixels, denoted as $\mathbb{B}_e$(effective pixels with the bright region), is defined as:
    \begin{equation}
        \mathbb{B}_e=\{ (x,y) \mid I_{EC}(x,y)\leq t_e(x,y)\}\cap \mathbb{B}, \label{con:11}
    \end{equation}
    here, \( t_e \) serves as a threshold determining a pixel's ability to adequately block light. It is inversely related to the derivative of the attenuation curve shown in Fig.~\ref{fig:effeciency}, as given by:
    \begin{equation}
        t_e(x,y) \propto \left(\dfrac{\mathrm{d}f_s}{\mathrm{d}I_m(x,y)} \right)^{-1}. \label{con:12}
    \end{equation}
    
    Conversely, the set of blurred pixels in the bright region is expressed as:
    \begin{equation}
        \mathbb{B}_b=\{ (x,y) \mid I_{EC}(x,y) > t_{e}(x,y)\}\cap \mathbb{B}. \label{con:13}
    \end{equation}
    Similarly, when effective and blurred pixels leak into the dark region, they can be defined as:
    \begin{equation}
        \mathbb{D}_e=\{ (x,y) \mid I_{EC}(x,y) \leq t_{e}(x,y)\}\cap \mathbb{D}. \label{con:14}
    \end{equation}
    \begin{equation}
        \mathbb{D}_b=\{ (x,y) \mid I_{EC}(x,y) > t_{e}(x,y)\}\cap \mathbb{D}. \label{con:15}
    \end{equation}
    
    It's important to note that sets \(\mathbb{B}\) and \(\mathbb{D}\) are fixed, as they are determined by the static scene under study. \(I_{EC}\), the image observed by the EC, changes if the occlusion mask is modified. We introduce the superscript \(\alpha\) on \(\mathbb{B}_e\) as \(\mathbb{B}^{\alpha}_e\) to denote its state when the expansion radius is \(\alpha\).
    
    The mask's expansion process is a morphological transformation, represented by:
    \begin{equation}
        I'_m(\alpha) = \min_{-\alpha<m,n<\alpha} I_m(x+m, y+n), \label{con:I'_m}
    \end{equation}
    where \( I'_m \) is the expanded occlusion mask.

    The optimization objective can be interpreted as maximizing the number of pixels in \(\mathbb{B}_e\) while minimizing the presence of pixels in \(\mathbb{D}_e\). Although pixels in \(\mathbb{B}_b\) can exist, their weight for either inclusion or exclusion becomes significant as they approach or deviate from the threshold \(t_e\). Pixels in \(\mathbb{D}_b\) follow a similar rule. Combining Eq. \eqref{con:10} through Eq. \eqref{con:I'_m}, the optimization problem can be formalized as:
    
    \begin{maxi}|s|
		{\alpha} {S(\alpha)=\sum_{(x,y) \in \mathbb{B}^{\alpha}_e} \sigma_{1}  + \sum_{(x,y) \in \mathbb{B}^{\alpha}_b} \sigma_{2} \cdot (1-2f_s(I_{EC}(x,y))}{}{}
		\breakObjective { \quad \quad \quad \quad - \sum_{(x,y) \in \mathbb{D}^{\alpha}_e} \sigma_{3} - \sum_{(x,y) \in \mathbb{D}^{\alpha}_b} \sigma_{4} \cdot f_s(I_{EC}(x,y))},  \label{con:16}
		\addConstraint{\alpha \in \mathbb{Z}^+}.
	\end{maxi}
    
    Here, \(\sigma_1\) to \(\sigma_4\) are the weights assigned to each pixel in the corresponding occlusion sets, and \(\mathbb{Z}^+\) denotes the set of positive integers.
    
    In terms of the weights, the contribution level of \(\mathbb{B}_e\) should be highest, and thus its weight should be set to the maximum. Meanwhile, the contribution or penalty level of the other three occlusion sets can be considered roughly equal. In practice, we find that \(\sigma_1\) should be set sufficiently high. Otherwise, Eq. \eqref{con:16} will lack an extremum. A unique extremum will manifest in Eq. \eqref{con:16} when \(\sigma_1\) is at least three times larger than \(\sigma_2\) through \(\sigma_4\), as demonstrated in Fig.~\ref{fig:opti-oofNopti}(b).
    
    Since \(\alpha\) represents discrete points, \(S(\alpha)\) is non-differentiable. However, given that this is a discrete optimization problem, the maximum value of \(S(\alpha)\) can still be efficiently found via brute-force search within a reasonable range.

	\section{Implementation}
    In this section, we describe the entire system implementation of our benchtop prototype. Our implementation covers the hardware and software setup, system calibration, and a critical algorithmic component—the out-of-focus PSF simulation. The out-of-focus simulation is integral for our optimization algorithm, allowing for controlled and precise calculations that mitigate the impacts of real-world aberrations such as noise and intensity distortions. This synergistic approach enables us to achieve high-precision, location-based selective dimming that is tailored to meet the unique needs of photophobic individuals.
    
	\subsection{Hardware and Software Setup} \label{sec41}
    As depicted in Fig.~\ref{fig:system}(bottom left), our prototype employs two identical monochrome cameras equipped with SONY IMX 287 sensors (720×540 pixels, 12-bit depth). The cameras feature a 6mm lens with an f/1.6 aperture, approximating a large baseline pupil diameter of 3.8mm~\cite{Anderson2009}. Arranged in parallel on a vertical axis, the cameras have a focal distance $D = 2$ meters from both the SC and the EC. They are connected via a USB 3.0 interface and transmit video data in real-time using a multi-threaded software development kit provided by the manufacturer. To minimize reflections, the entire hardware setup is covered with a black curtain.

    The primary and secondary LCDs consist of SONY LCX017 panels (1024×768 pixels, 36.9×27.6mm active area, 705 dpi, 60 Hz, monochrome). Post-power-on transmittance for these panels is 21\%, dropping to 10\% when the linear polarizer is applied. The distance from each camera to its corresponding LCD panel is $D'=10$mm. The LCD panels are managed by an external driver board connected to a computer via a VGA link.

    Our software, implemented in Python 3.8, utilizes OpenCV-Python for system calibration and image processing. The computer configuration comprises an AMD 8-core 3.8GHz CPU and an NVIDIA GTX Titan X GPU.
    
    The framework of the prototype is 3D-printed in plastic, designed using Autodesk\textregistered Fusion 360 and printed by ANYCUBIC.
    
	\subsection{System Calibration}
	\subsubsection{Calibration between the EC and the primary LCD}
    As shown in Fig.~\ref{fig:layout}, the EC-to-LCD mapping adheres to a perspective model, described by a homography matrix. We adjusted the EC focus alternately between the focal and LCD planes, recording the corresponding projections. Using OpenCV-Python, we calculated the homography matrix based on these observations.
 
	\subsubsection{Calibration between the EC and the SC}
    Similarly, the mapping between the SC and the EC is described by a 2D homography matrix. We acquired images of an A1 size chessboard located on the focal plane using both the SC and EC, and calculated the homography matrix using OpenCV-Python functions.

    \begin{figure} [t]
    \centering
    \includegraphics[width=0.6\textwidth]{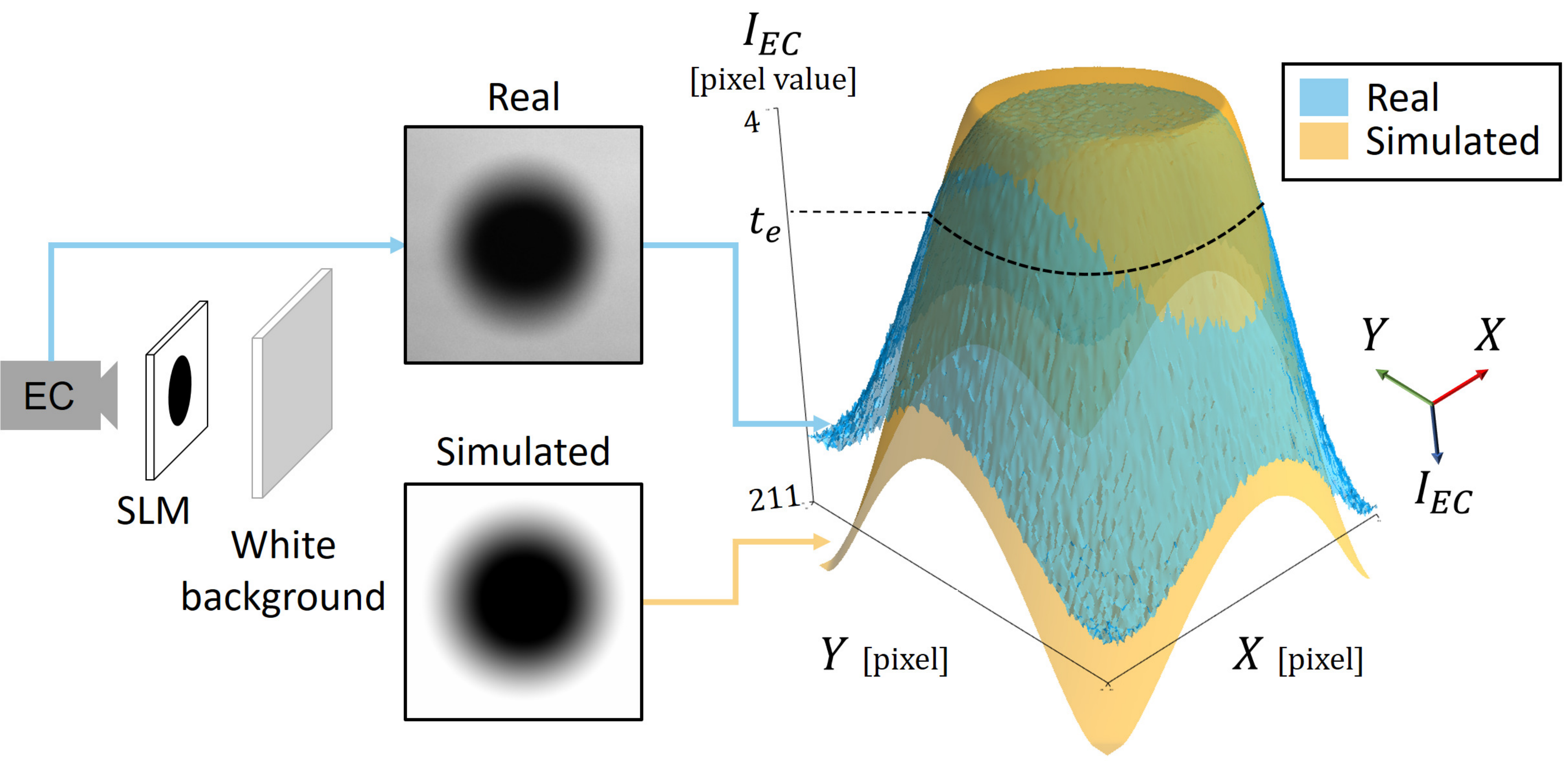}
    \caption{
        (Left) The EC captures the real occlusion by positioning a white background in front of the SLM. (Middle) Result of the out-of-focus simulated occlusion mask. (Right) 3-D visualization comparing it with the real occlusion.}
    \label{fig:simulate}
    \end{figure}

    \subsection{Out-of-focus PSF Simulation} \label{sec51}
     As shown in Fig.\ref{fig:simulate}, we start by displaying a static circle-shaped occlusion mask (with a 100-pixel radius) on the primary LCD. To capture a clean occlusion mask, the EC records the mask while a white background is in place before the benchtop system. Background subtraction yields the pure circle mask. Based on Eq. \eqref{con:7} and \eqref{con:8}, we formulate the problem as:  
    \[\arg \min_r \lVert I_m \otimes H_{OOF}(r) - I_{EC} \rVert_2.\]
    Here, the Euclidean distance is employed to quantify the similarity. To mitigate the effects of noise and diffraction aberrations originating from edge pixels, we restrict the effective calculation area to the central portion of the circle mask. It is worth noting that the simulated occlusion mask does not need to be a perfect match with the real mask, which could be distorted due to noise or diffraction. Because the primary function of the simulated mask is to serve as a performance metric, we consider the \textit{effective occlusion} area of the simulated mask. Thus, the final simulated mask should ensure that its \textit{effective occlusion} area closely aligns with that of the real mask, as demonstrated in Fig.~\ref{fig:simulate}(right). The areas of the real and simulated 3-D figures within \(t_e\) are virtually identical.

	\section{Experiments and Results}
    In this section, we demonstrate the effectiveness of our control and optimization algorithms through practical experiments conducted with our proof-of-concept system. Section~\ref{sec52} details the comparative studies of different masking techniques, including the \naive mask, aperture-based expansion mask, and our optimized mask. Section~\ref{sec53} discusses how performance varies with calibration deviations and evaluates the utility of additional blurring processes.
        	
        \begin{figure}
            \centering
            \subfigure[]{
                \includegraphics[width=0.33\textwidth]{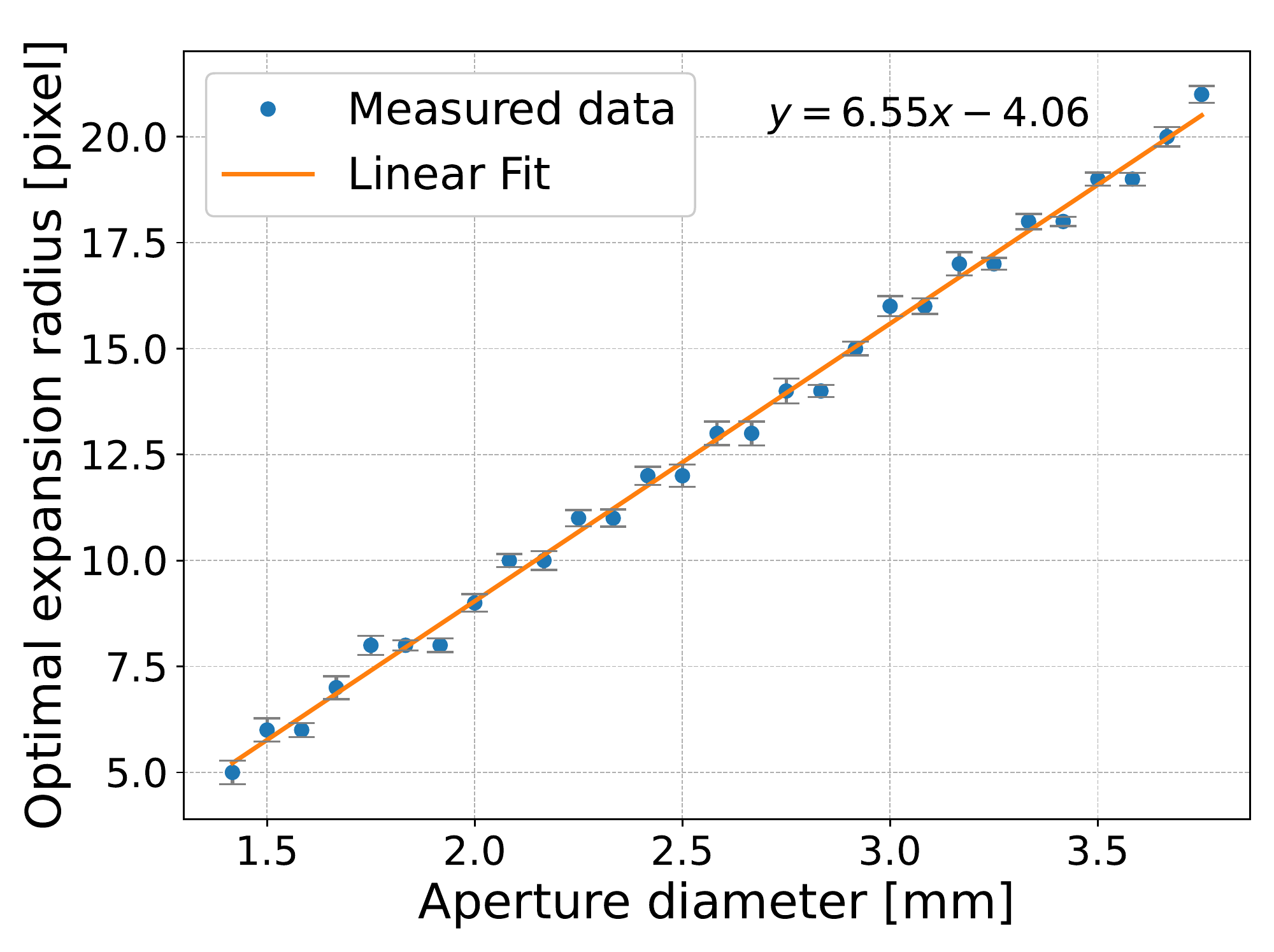}
            }
            \subfigure[]{
                \includegraphics[width=0.32\textwidth]{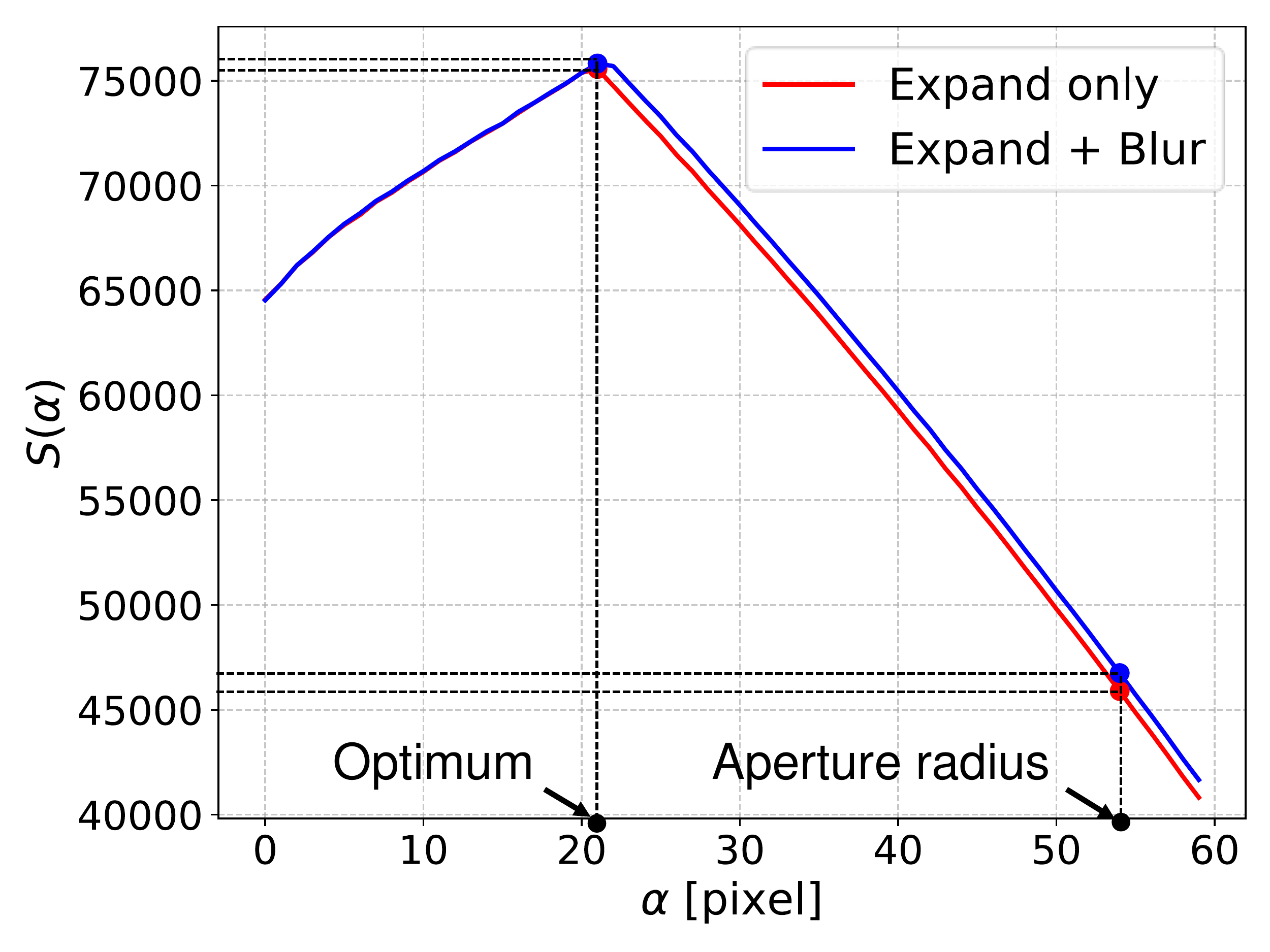}
            }
            \caption{
                (a) Depicting the mapping relationship between the optimal expansion radius and the aperture diameter, enabling the system to compute an optimized mask from the user's pupil size.
                (b) Visualization of Eq. \eqref{con:16} with an optimum (extremum) at \( \alpha=21 \). Our optimization model demonstrates that the function \( S(\alpha) \) computed using the aperture radius is notably lower than the one derived using the optimal radius. The blue curve represents \( S(\alpha) \) when an additional blurring process is incorporated. Although the optimum of \( S(\alpha) \) with added blurring is greater than that of a simple expansion, the difference is not significant.
            }
            \label{fig:opti-oofNopti}
        \end{figure}

	\begin{figure}[t]
		\centering
		\includegraphics[width=0.7\linewidth]{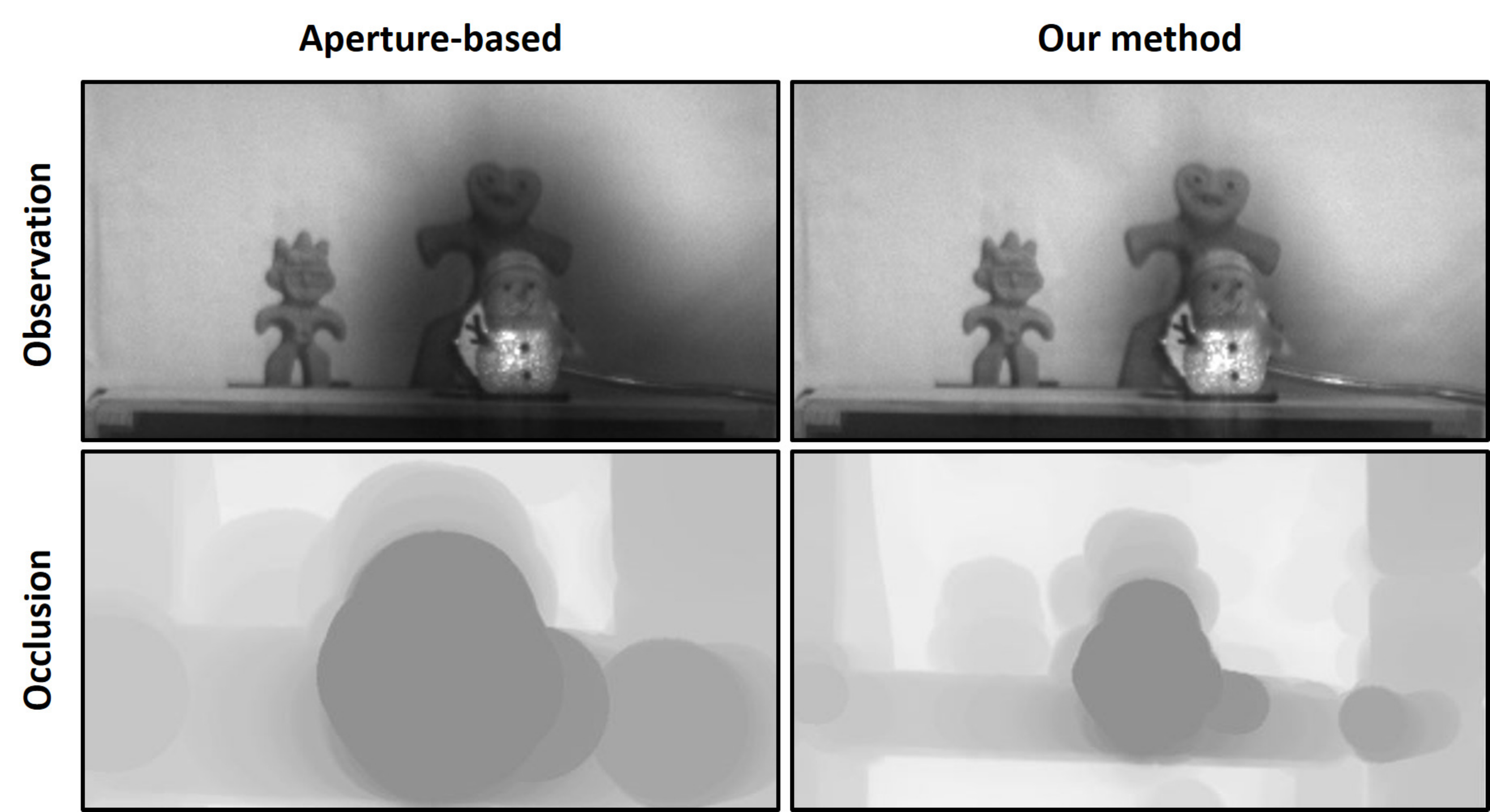}
		\caption{
			Visualization of occlusion leak. By presenting a static occlusion mask on the LCD, the leak becomes pronounced when the background adopts a lighter hue.
		}
		\label{fig:result-aperture-method}
	\end{figure}

\begin{figure}[!t]
\centering
\includegraphics[width=0.9\linewidth]{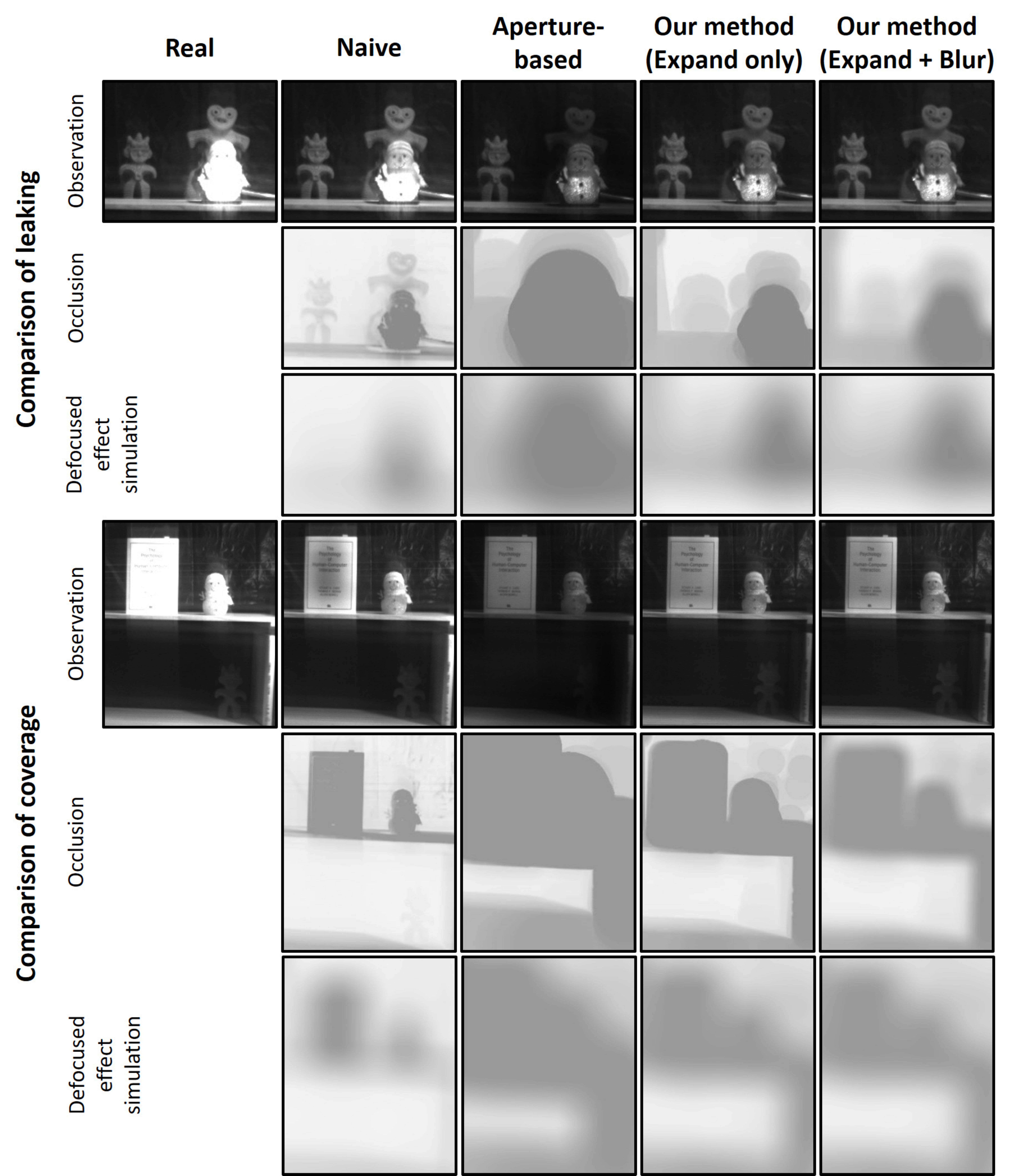}
\caption{Real-time observation results for the \naive mask, aperture-based expansion mask, and our proposed masks. Beneath each mask is its corresponding out-of-focus simulated effect.
}
\label{fig:results}
\end{figure}

	\subsection{Control and Optimization Algorithm} \label{sec52}
    Our system employs a two-step approach that utilizes both control and optimization algorithms for improved performance. Initially, the control algorithm generates a \naive mask using the modulation function (shown in Fig.~\ref{fig:function}) to adjust the pixel intensity, particularly aiming to suit the needs of photophobic individuals.

    Following this, our optimization algorithm refines the \naive mask to mitigate occlusion leak and balance image contrast, particularly in high-contrast settings. This refinement is essential for preventing unwanted brightness and improving the visual experience, a critical need when aiming to minimize eye strain for photophobic individuals.

    Utilizing the out-of-focus PSF simulated during the implementation phase, we compared the effectiveness of our method against \naive and aperture-based expansion masks. Here, the \naive mask is generated using the control algorithm but without subsequent optimization. The aperture-based expansion radius \(\alpha_a\) is calculated as follows:
    \[
	\alpha_a = \frac{a}{2}\cdot \frac{36.9}{1024} \approx 53.
	\]
     In contrast, our optimization algorithm suggests an optimal expansion radius of \(\alpha=21\). The mapping between this optimal radius and the aperture diameter was initially derived through simulation and further validated empirically, as illustrated in Fig.~\ref{fig:opti-oofNopti}(a).

    For the actual experiments, we set the camera aperture size to 3.8 mm and adjusted the exposure settings to simulate photophobic vision. The outcomes are detailed in Fig.~\ref{fig:results}. To assess the algorithm's ability to reduce occlusion leak, we used a translucent snowman doll illuminated by an incandescent bulb placed behind it. Additional clay statues situated behind the bulb were lit only by ambient room light. Using the \naive mask, the snowman appeared overly bright due to the mask's limited blocking capabilities, especially when the object is out-of-focus. Conversely, the aperture-based mask, when expanded to block all incoming light, resulted in a significant occlusion leak and unintentional masking of the clay statues. The occlusion leak worsened as the backdrop color lightened, as shown in Fig.~\ref{fig:result-aperture-method}. 

    Our results conclusively demonstrate that our image-level optimization algorithm minimizes occlusion leak while maintaining satisfactory image contrast. Further tests involving a reference object with well-defined edges indicated that our algorithm excels at masking, even in conditions featuring blurry edges.

   


\subsection{Additional Blurring Process} \label{sec53}

We further investigated the potential benefits of adding a blurring process after the mask expansion. The goal of this additional step is to soften the edges of the mask, thereby enhancing the overall masking effect. According to results obtained from our optimization algorithm, the addition of this blurring process yields higher maximum intensity values compared to simple mask expansion alone, as illustrated by the blue curve in Fig.\ref{fig:opti-oofNopti}(b). However, the advantages of this additional process are not overwhelmingly significant, as evident from the rightmost column in Fig.\ref{fig:results}.

\begin{figure}[!t]
		\centering
		\includegraphics[width=0.65\linewidth]{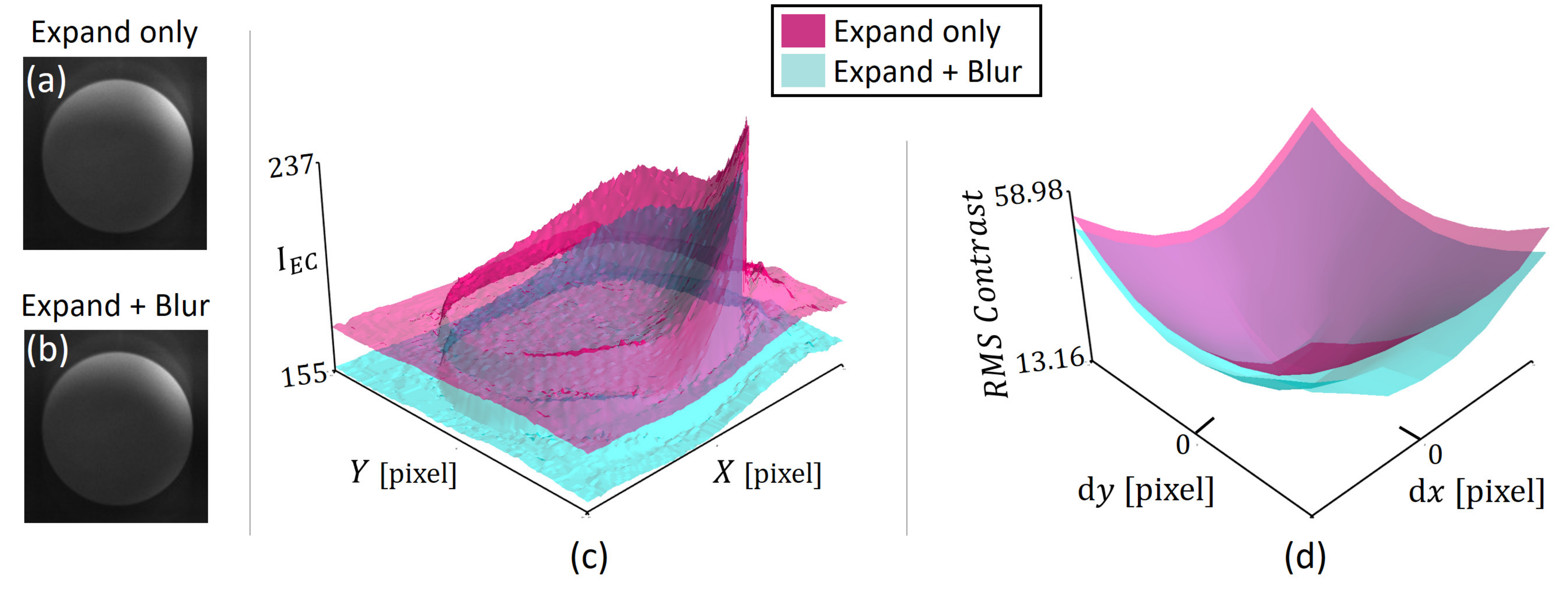}
		\caption{
			(a), (b) Experiments assessing calibration deviation using two proposed masks. (c) Surface plot illustrating the corresponding occlusion effect. (d) RMS contrast comparison between the two proposed methods, evaluated across various calibration deviations. The X-axis and Y-axis represent the directions of deviation.
		}
		\label{fig:aberration}
	\end{figure}

Subsequently, we compared the surface plots generated with and without the blurring process. Accounting for the typical calibration deviations that may occur in everyday use, we found that a mask with blurred edges is more effective at blocking incident light at the peripheries, as depicted in Fig.\ref{fig:aberration}. Further calculations on the root mean square (RMS) contrast indicated that the added blurring process also somewhat reduces scene contrast, as shown in Fig.\ref{fig:aberration}(d).

\section{Conclusions and Future Work}
    We presented a smart dimming sunglasses system designed to alleviate photophobic discomfort. Our solution leverages SLM technology and a monochrome camera for real-time scene adaptation. Unlike traditional sunglasses and existing global dimming eyewear that have limitations in complex lighting scenarios, our system provides flexible, location-based light modulation. Our control and optimization algorithms allow for tailored visual experiences, enabling users to comfortably navigate varying lighting conditions without compromising visibility. Experimental results demonstrated the efficacy of our optimized masks in achieving more effective light modulation compared to \naive and aperture-based expansion techniques.
  
    In future work, we aim to resolve the system's current limitations. Upgrading the LCD panel's light transmittance is a priority, with the advent of polarizer-free liquid crystal dimmers~\cite{Talukder2019} offering a promising avenue. Incorporating eye-tracking technology can fine-tune the mask modulation, as it is directly related to the pupil size of the user. To improve computational speed, particularly for the convolution operations needed for mask optimization, we consider the integration of specialized hardware such as field-programmable gate arrays (FPGAs) or application-specific integrated circuits (ASICs). Incorporating such hardware would not only make the system more efficient and responsive but also pave the way for a truly wireless, everyday wearable dimming sunglasses, elevating the overall user experience.

      \bibliographystyle{plainnat}

\begin{thebibliography}{99}

\bibitem{AlphaMicron2023}
Inc AlphaMicron.
\newblock Ctrl eyewear, 2023.
\newblock Accessed: 2023-09-09.

\bibitem{Anderson2009}
Christa~J. Anderson and John Colombo.
\newblock {Larger tonic pupil size in young children with autism spectrum
  disorder}.
\newblock {\em Developmental Psychobiology}, 51(2):207--211, 2009.

\bibitem{armistead1964photochromic}
WH~Armistead and SD~Stookey.
\newblock Photochromic silicate glasses sensitized by silver halides: Their
  characteristic of changing color reversibly, in combination with other
  properties, suggests many uses.
\newblock {\em Science}, 144(3615):150--154, 1964.

\bibitem{bhagavathula2007extremely}
Kiriti Bhagavathula, Albert~H Titus, and Christopher~S Mullin.
\newblock An extremely low-power cmos glare sensor.
\newblock {\em IEEE Sensors Journal}, 7(8):1145--1151, 2007.

\bibitem{burstein2019neurobiology}
Rami Burstein, Rodrigo Noseda, and Anne~B Fulton.
\newblock The neurobiology of photophobia.
\newblock {\em Journal of neuro-ophthalmology: the official journal of the
  North American Neuro-Ophthalmology Society}, 39(1):94, 2019.

\bibitem{chandrasekhar2014matched}
Prasanna Chandrasekhar, Brian~J Zay, Chunming Cai, Yanjie Chai, and David
  Lawrence.
\newblock Matched-dual-polymer electrochromic lenses, using new cathodically
  coloring conducting polymers, with exceptional performance and incorporated
  into automated sunglasses.
\newblock {\em Journal of Applied Polymer Science}, 131(22), 2014.

\bibitem{Clark2017}
Joseph Clark, Kimberly Hasselfeld, Kathryn Bigsby, and Jon Divine.
\newblock {Colored Glasses to Mitigate Photophobia Symptoms Posttraumatic Brain
  Injury}.
\newblock {\em Journal of Athletic Training}, 52(8):725--729, 2017.

\bibitem{deb1969novel}
SK~Deb.
\newblock A novel electrophotographic system.
\newblock {\em Applied Optics}, 8(101):192--195, 1969.

\bibitem{dumas2012fast}
JC~Dumas, J~Vidal, and V~Dumas.
\newblock Fast response liquid crystal glasses.
\newblock {\em Lighting Research \& Technology}, 44(4):498--505, 2012.

\bibitem{Fan2009}
Xiaofei Fan, Judith~H. Miles, Nicole Takahashi, and Gang Yao.
\newblock {Abnormal transient pupillary light reflex in individuals with autism
  spectrum disorders}.
\newblock {\em Journal of Autism and Developmental Disorders},
  39(11):1499--1508, 2009.

\bibitem{good1991use}
PA~Good, RH~Taylor, and MJ~Mortimer.
\newblock The use of tinted glasses in childhood migraine.
\newblock {\em Headache: The Journal of Head and Face Pain}, 31(8):533--536,
  1991.

\bibitem{goodman2005introduction}
Joseph~W Goodman.
\newblock {\em Introduction to Fourier optics}.
\newblock Roberts and Company publishers, 2005.

\bibitem{hainich2016displays}
Rolf~R Hainich and Oliver Bimber.
\newblock {\em Displays: fundamentals \& applications}.
\newblock CRC press, 2016.

\bibitem{hiroi2017adaptivisor}
Yuichi Hiroi, Yuta Itoh, Takumi Hamasaki, and Maki Sugimoto.
\newblock Adaptivisor: Assisting eye adaptation via occlusive optical
  see-through head-mounted displays.
\newblock In {\em Proceedings of the 8th Augmented Human International
  Conference}, pages 1--9, 2017.

\bibitem{Itoh2017}
Yuta Itoh, Takumi Hamasaki, and Maki Sugimoto.
\newblock {Occlusion Leak Compensation for Optical See-Through Displays Using a
  Single-Layer Transmissive Spatial Light Modulator}.
\newblock {\em IEEE Transactions on Visualization and Computer Graphics},
  23(11):2463--2473, 2017.

\bibitem{katz2016diagnosis}
Bradley~J Katz and Kathleen~B Digre.
\newblock Diagnosis, pathophysiology, and treatment of photophobia.
\newblock {\em Survey of ophthalmology}, 61(4):466--477, 2016.

\bibitem{1240696}
K.~Kiyokawa, M.~Billinghurst, B.~Campbell, and E.~Woods.
\newblock An occlusion capable optical see-through head mount display for
  supporting co-located collaboration.
\newblock In {\em Proceedings of the Second IEEE and ACM International
  Symposium on Mixed and Augmented Reality, 2003}, pages 133--141, 2003.

\bibitem{lebensohn1934nature}
JAMES~E LEBENSOHN and John Bellows.
\newblock The nature of photophobia.
\newblock {\em Archives of Ophthalmology}, 12(3):380--390, 1934.

\bibitem{lee20203d}
Joong~Hoon Lee, Hanseop Kim, Ji-Young Hwang, Jinmook Chung, Tae-Min Jang,
  Dong~Gyu Seo, Yuyan Gao, Junhyun Lee, Haedong Park, Seungwoo Lee, et~al.
\newblock 3d printed, customizable, and multifunctional smart electronic
  eyeglasses for wearable healthcare systems and human--machine interfaces.
\newblock {\em ACS applied materials \& interfaces}, 12(19):21424--21432, 2020.

\bibitem{ma2008smart}
Chao Ma, Minoru Taya, and Chunye Xu.
\newblock Smart sunglasses based on electrochromic polymers.
\newblock {\em Polymer Engineering \& Science}, 48(11):2224--2228, 2008.

\bibitem{main2000wavelength}
Alan Main, Ioannis Vlachonikolis, and Andrew Dowson.
\newblock The wavelength of light causing photophobia in migraine and
  tension-type headache between attacks.
\newblock {\em Headache: The Journal of Head and Face Pain}, 40(3):194--199,
  2000.

\bibitem{Nayar2003}
Shree~K. Nayar and Vlad Branzoi.
\newblock {Adaptive dynamic range imaging: Optical control of pixel exposures
  over space and time}.
\newblock In {\em Proceedings of the IEEE International Conference on Computer
  Vision}, volume~2, pages 1168--1175, 2003.

\bibitem{osterby1991photochromic}
Bruce Osterby, Ronald~D McKelvey, and Lisa Hill.
\newblock Photochromic sunglasses: A patent-based advanced organic synthesis
  project and demonstration.
\newblock {\em Journal of Chemical Education}, 68(5):424, 1991.

\bibitem{osterholm2015four}
Anna~M \"Osterholm, D~Eric Shen, Justin~A Kerszulis, Rayford~H Bulloch, Michael
  Kuepfert, Aubrey~L Dyer, and John~R Reynolds.
\newblock Four shades of brown: tuning of electrochromic polymer blends toward
  high-contrast eyewear.
\newblock {\em ACS applied materials \& interfaces}, 7(3):1413--1421, 2015.

\bibitem{rao2022low}
Tingke Rao, Yuanliang Zhou, Jie Jiang, Peng Yang, and Wugang Liao.
\newblock Low dimensional transition metal oxide towards advanced
  electrochromic devices.
\newblock {\em Nano Energy}, 100:107479, 2022.

\bibitem{Talukder2019}
Javed~Rouf Talukder, Hung-Yuan Lin, and Shin-Tson Wu.
\newblock {Photo- and electrical-responsive liquid crystal smart dimmer for
  augmented reality displays}.
\newblock {\em Optics Express}, 27(13):18169, 2019.

\bibitem{vincent1989controlled}
Arnaud~JP Vincent, Egilius~LH Spierings, and Harley~B Messinger.
\newblock A controlled study of visual symptoms and eye strain factors in
  chronic headache.
\newblock {\em Headache: The Journal of Head and Face Pain}, 29(8):523--527,
  1989.

\bibitem{Wetzstein2010}
Gordon Wetzstein, Wolfgang Heidrich, and David Luebke.
\newblock {Optical image processing using light modulation displays}.
\newblock {\em Computer Graphics Forum}, 29(6):1934--1944, 2010.

\bibitem{yang2021highly}
Zetian Yang, Jiaren Du, Lisa~IDJ Martin, David Van~der Heggen, and Dirk
  Poelman.
\newblock Highly responsive photochromic ceramics for high-contrast rewritable
  information displays.
\newblock {\em Laser \& Photonics Reviews}, 15(4):2000525, 2021.

\end{thebibliography}


\end{document}